\documentclass{jpsj3}

\usepackage{graphicx}%
\usepackage{dcolumn}%
\usepackage{bm}%

\title{Melting of MgO Studied using a Multicanonical Ensemble Method Combined with a First-Principles Calculation}%

\author{Yoshihide Yoshimoto
\thanks{E-mail address: yosimoto@issp.u-tokyo.ac.jp}
}
\inst{%
Institute for Solid State Physics, University of Tokyo,
5-1-5 Kashiwa-no-ha, Kashiwa-shi, Chiba, 2778581, Japan
}

\abst{
Melting of MgO was studied using a multicanonical ensemble method
combined with a first-principles calculation.
This approach has been successively performed by using a rather simple functional form for a
model inter-atomic potential that is determined from first-principles and a
novel approximation treating auxiliary degrees of freedom,
such as electron thermal excitations, within a multicanonical ensemble
method.
Although a rather simple model potential was used,
this approach could distinguish differences due to
the exchange-correlation potential used in the first-principles calculations.
The pressure dependence of the melting point,
latent heat, and volume change during melting were studied.
The obtained dependence was similar to that reported by Alf\`e
which differs from experimental results.
This dependence did not change even with
the PBEsol exchange-correlation potential.
}

\kword{%
MgO, multicanonical ensemble, first-principles calculation, melting, pressure dependence}

\begin{document}
\maketitle

\section{Introduction}
Changes in atomic structure during phase transitions has been one of the principal
goals in material science. Of the various types of phase transitions, melting,
and its reverse process, crystallization, are the most basic.
Moreover, these are both technologically important because
crystal growth is fundamental to material synthesis as
is casting to shape forming.

In another aspect, first-principles calculations has been
a powerful theoretical tool in material science because of its ability to
treat atomic structures. This ability is one reason why
it is expected to contribute future material developments.

In general, material developments can be separated into a cycle of three
stages: design, synthesis, and characterization.
The synthesis stage is, however, one of the less-developed areas as far as
first-principles studies are concerned.
Most first-principles studies have contributed to the characterization of materials.
 With the exception of some advanced work, few have been proposed in the design stage.
For this reason, the present study endeavors to treat melting by
a first-principles calculation. First-principles treatment of crystal growth and melting,
two processes that are strongly related within synthesis work, should become common practice.

In the study of melting employing first-principles calculations,
two approaches exist.
The first is the two-phase method, which simulates an interface between
two phases to determine melting points ($T_m$)\cite{MWHC94}.
The other is the thermodynamic integration method\cite{FS02chap7}
and related adiabatic switching method.\cite{WR90,SC95}
The first requires a large simulation cell to simulate adequately
an interface between two phases.
The other does not require such a large simulation cell but requires
some reference system whose free energy is well-understood and can be
smoothly connected to the target system under study.

The author has previously proposed another first-principles approach to the study of melting\cite{Yoshimoto06},
based on a combination of the multicanonical ensemble method\cite{BN91,BN92,Lee93} along with
a first-principles calculation.
It does not require a large simulation cell because it does not
simulate an interface. This is important from a first-principles
calculation perspective because its computational cost increases rapidly
as a function of system size.
Furthermore, the approach does not require a common well-understood reference system 
connecting the free energies of both phases.
In general, such systems are not expected to exist.

The present study treats the melting of MgO by this novel approach.
The reason why MgO was chosen is as follows.
(1) Of the four types of crystals, covalent, metallic, ionic, and molecular,
MgO is a typical representative of ionic crystals.
(2) MgO has technological applications. It is a typical
refractory material and a typical substrate used in thin film growth.
(3) MgO is one extreme constituent of rocks, which are oxides of
Mg, Si, and Al. For this reason, the pressure ($P$) dependent
melting curve of MgO has been studied extensively in the context of earth science.
An experimental study has suggested that MgO melts at rather low temperature
under high pressures.\cite{ZB94} (Specifically, Zerr and Boehler observed that MgO melted at
$\sim 3500$ K under $\sim 17$ GPa, and predicted
that MgO melts at $\sim 5200$ K under $\sim 135$ GPa.)
However, a following first-principles study using the two phase method
did not agree with it. \cite{Alfe05}
(i.e., at $4590$ K and $8144$ K under $17$ GPa and $135.6$ GPa, respectively)
In addition, another experimental study by
Zhang and Fei on (Mg,Fe)O solidus under
high pressure recently claimed higher $T_m$ than
theoretical values under high pressures from
the extrapolation of their result.\cite{ZF08} (Even under $7$ GPa,
the estimated $T_m$ was $\sim 4500$ K.
The extrapolated melting point from their result
under $17$ GPa seems to be $\sim 5500$ K.)
Altanetively, we can compare the melting slope ($dT_m/dP$)
instead of $T_m$ itself, because
we can calculate $dT_m/dP$ from the Clausius-Clapeyron relation.
Recently, Tangeny and Scandolo\cite{TS09} studied
this melting slope at $P=0$ GPa using a combination of
molecular dynamics and density functional calculations.
After a detailed discussion, they
concluded that the theoretical values are from $\sim 130$ to
$\sim 150$ K/GPa, which clearly differs from
the ones reporeted by Zerr and Boehler ($36$ K/GPa).
(4) The melting behavior of MgO is not fully understood. No calorimetric
study of latent heat is available for this material. The apparent values ranging from 8
to 30 kcal mol$^{-1}$ were obtained from binary systems.\cite{CDDFMS85}

The present paper also reports the application of
the PBEsol exchange-correlation (XC) potential\cite{PRCVSCZB08,PRCVSCZB09}
to investigate a possible cause for the discrepancy
between experimental and theoretical results.
This issue may be because of an electronic correlation
problem in first-principles calculations and the PBEsol claim that
it is a better approximation for condensed materials.

The structure of the paper is as follows:
First, a brief introduction to the present approach is presented.
Second, I address an issue arising with thermal excitation of electrons and its treatment.
Third, the definition of an order parameter is discussed within the present approach.
Fourth, calculation conditions will be detailed.
Fifth, calculational checks are presented.
Sixth, comparisons between various XC potentials used in the
present approach will be analyzed.
Last, the simulated melting behavior at $P=0$GPa and the pressure dependence
of melting will be presented for the issues concerning latent heat and 
melting point curves.

The simulated results will yield values for $T_m$,
latent heat ($\Delta H$), and volume change per atom during melting ($\Delta V$).
These properties have significance in material synthesis.
For instance, the melting point and its latent heat will be useful in planning
crystal growth. The melting point itself becomes important when
developing heat-resistant materials.
Volume changes during melting will be useful information for
casting when precision is important in material shaping. Pressure effects
are relevant in high-pressure synthesis work.

\section{Multicanonical ensemble method combined with first-principles calculation}

In this section, a brief review is given highlighting the combination of a multicanonical ensemble method
with first-principles calculations.

The multicanonical ensemble method is a type of generalized ensemble
method that uses a generalized statistical weight instead of the canonical
Boltzmann factor $e^{-\beta E}$. This generalized statistical weight is the estimated inverse of the density of states $\tilde{W}(E)$ for which the probability $P(E)$ of
observing some energy $E$
becomes constant. This is because $P(E) \propto W(E)\tilde{W}^{-1}(E) \sim 1$, where, $W(E)$
is the density of states in the system. As a consequence, the energy observed in a molecular dynamics (MD) simulation
or a Monte Carlo simulation fluctuates widely so that the tunneling time
from one state to another is expected to be short
compared with that in a canonical simulation.
There, the fluctuation is strongly
confined around the expected energy value.
$\tilde{W}$ or the estimated entropy $\tilde{S} = k_B\log\tilde{W}$
can be generated by the Wang-Landau algorithm.\cite{WL01a} This is
an ``on-the-fly''-type algorithm that is used to obtain $\tilde{S}$ in an efficient manner.

To obtain any physical quantity from a multicanonical simulation,
a technique to recompile the simulation run, called re-weighting,\cite{FS88}
is used. Unlike in a canonical simulation, a physical quantity
at {\em any} temperature can be obtained ideally from just a 
{\em single} simulation run by re-weighting.

The multicanonical ensemble method works theoretically for a system with a
first-order phase transition. However, naively applying this method to
simulate melting of a crystal such as silicon or MgO is still difficult
because of the slow tunneling times between the two phases.

To overcome this difficulty, we found in a previous study that the multi-order multi-thermal
({\em MOMT}) ensemble\cite{WL01b}, utilizing an order parameter
defined using structure factors of the target crystal structure, was
useful.\cite{Yoshimoto06}
This ensemble is a two-component multicanonical
ensemble.\cite{HNSKN97}
Specifically, $\tilde{S}$ becomes a function of both energy and an
order parameter. By explicitly treating the crystalline order,
we can make the system fluctuate between two phases in short periods.

This extended multicanonical ensemble method can in practice simulate the
phase transition with a model inter-atomic potential.
Direct application with first-principles calculations is, however,
still impractical because the required number of MD
steps is far larger than available number ($\sim 10^4$).

Thermodynamic downfolding is a method that resolves this issue between available number and required number of the steps.
This method generates a model inter-atomic potential $U_M$
from an accurate inter-atomic potential $U_A$
(for instance, one obtained from first-principles) conserving the thermodynamics
of the target system as much as possible.

Usually, a model inter-atomic potential is constructed
by making it reproduce as much as possible an accurate version on
some reference atomic configurations.
In thermodynamic downfolding, the reference configurations
are a (down-sampled) multicanonical ensemble (simulation run).
This choice is thermodynamically meaningful because we can
derive any thermal quantity at any temperature by applying
a re-weighting over the ensemble.
Specifically, a multicanonical ensemble is a representative of the 
total thermodynamics of a system. Dependence on the chosen atomic configurations is eliminated
by making the inter-atomic potential and the configurations
self-consistent.

In summary, the combination of the {\em MOMT} ensemble method
and the thermodynamic downfolding makes multicanonical simulations
of melting from first-principles possible.
The merits of this approach are as follows:
(1) The required simulation cell size
is small because we do not simulate an interface between the two phases.
(2) We do not have to provide a reference inter-atomic potential
for which free energy is well-understood and which can be
smoothly connected to the target potential.
To provide such a potential is difficult especially for a system with
multiple components.
(3) Accurate calculations can be performed in
parallel because all atomic configurations to be calculated
are known from the outset. Therefore, we can utilize inexpensive computer systems
with weak interconnects.
In addition, this approach does not need any parameter-fitting
of some experimental data. Indeed, all parameters in
the model inter-atomic potential are determined from first-principles.
In this context, this approach represents a type of {\em ab}-{\em initio}
method. The form of the model potential should be regarded as 
a kind of basis set. In this approach, we can chose it arbitrarily so that the residual in the downfolding
becomes small enough.

\section{Approximation to treat auxiliary degrees of freedom such as thermal excitation of electrons}

The rather high melting point of MgO makes the thermal excitation of
electrons have a significant effect on the melting.\cite{Alfe05}
However, in trying to include such effects, the multicanonical ensemble method has a problem treating auxiliary degrees of freedom,
such as thermal excitation of electrons to be ¡Ètraced out¡É.
In contrast, we can partially trace out
the auxiliary degrees of freedom in the canonical ensemble method.
This partial trace gives a free energy function as a function of both
the principal degrees of freedom (e.g. the atomic positions)
and temperature. By using the free energy
function instead of the energy function in a Monte-Carlo simulation or
a MD simulation, we can deduce a thermal average of
a physical quantity that does not directly depend on the auxiliary
degrees of freedom.
In a multicanonical simulation, however, the energy function
itself cannot depend on temperature because the simulation run
does not have a physical temperature.

To resolve this issue, an approximation is developed.
To understand the approximation, the re-weighting method employed
to obtain the average of physical quantity should be reconsidered.
This thermal-averaged physical quantity $A$ is calculated by
a re-weighting method from the expression
\begin{equation}
\langle A\rangle = \frac{\sum_{i} A_i \tilde{W}(E_i)e^{-\beta E_i}}{\sum_i \tilde{W}(E_i)e^{-\beta E_i}},
\end{equation}
where $E_i$ is the energy of a configuration $i$ in a multicanonical
simulation run.

\begin{figure}
  \begin{center}
  \includegraphics{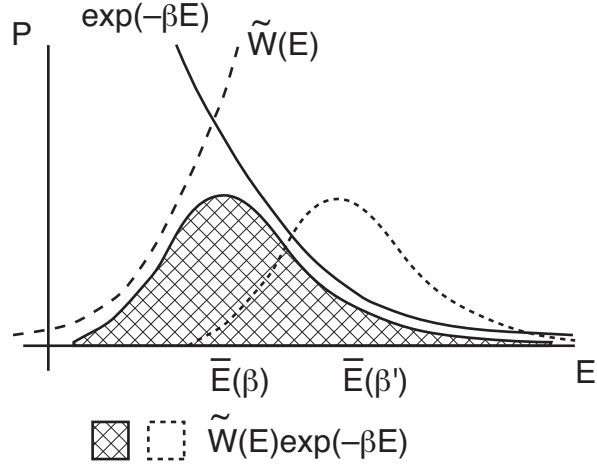}
  \end{center}
  \caption{\label{fig:rewgtdetail} The term, $\tilde{W}(E)e^{-\beta E}$, for the multicanonical ensemble to obtain physical quantity. $E$ and $P$ are energy and probability, respectively. $\tilde{W}(E)$ is an increasing function, while $e^{-\beta E}$ is a decreasing function. The product, $\tilde{W}(E)e^{-\beta E}$, has a peak at $\bar{E}$ and its position depends on temperature.}
\end{figure}

Here, we note the following fact: the term $\tilde{W}(E)e^{-\beta E}$ has
a sharp peak around the expectation value of energy $\bar{E}(\beta)$ (Fig. \ref{fig:rewgtdetail}).
We can consider this term as a ¡Èclipping mask¡É for the multicanonical
simulation run. Because of this fact, each
configuration $i$ has its maximal contributing temperature $\beta_{i}^{eff}$, obtained by finding the temperature that maximizes the following expression
\begin{equation}
\frac{\tilde{W}(E_i)e^{-\beta E}}{\sum_i \tilde{W}(E_i)e^{-\beta E_i}}.
\end{equation}

By using $\beta_{i}^{eff}$, we can determine a model inter-atomic potential
by thermodynamic downfolding as follows:
a first-principles calculation is performed at temperature
$\beta_{i}^{eff}$ on each configuration $i$
in the down-sampled multicanonical ensemble
. In particular, by using $\beta_{i}^{eff}$, we can
incorporate approximately the temperature dependence of each $i$.
The first-principles calculation produces both energy and
free energy. For both, thermodynamic downfolding is performed
to obtain both the model inter-atomic energy and 
free energy potentials.
The following multicanonical simulation is performed with this
inter-atomic free energy potential because the statistical weight is
determined by the free energy. To obtain the thermal-averaged energy,
however, we have to average the inter-atomic energy potential
instead of the free energy potential because in the partial trace formalism we have
\begin{equation}
\bar{E}(\beta) = \frac{\sum_i E_i e^{-\beta F_i(\beta)}}{\sum_i e^{-\beta F_i(\beta)}}.
\end{equation}

Consequently, in the present approximation, the re-weighting method for energy becomes

\begin{equation}
\bar{E}(\beta) = \frac{\sum_i E_i \tilde{W}(F_i)e^{-\beta F_i}}{\sum_i \tilde{W}(F_i)e^{-\beta F_i}}.
\end{equation}

In applications to melting, a further improvement is available.
We can split the down-sampled multicanonical ensemble into
crystalline-like and liquid-like configurations
by using an order parameter. Therefore, it is reasonable to
use the split ensemble including the target configuration $i$
to determine $\beta_i^{eff}$. By this refinement, we can
better treat over-heated states and over-cooled states.

\section{The order parameter for multi-order multi-thermal simulation for MgO}

To perform a multi-order multi-thermal simulation efficiently,
a rescaled order parameter is introduced:
\begin{equation}
O = \sinh\left(\frac{O_{ns}-\frac{1}{2}O_{max}}{\sqrt{O_{max}/O_{\alpha}}}\right)/\sinh\left(\frac{1}{2}\sqrt{O_{\alpha}O_{max}}\right) + \frac{O_{ns} - \frac{1}{2}O_{max}}{O_{max}},\ \textrm{where}\ O_{ns} = \frac{1}{N_\mathcal{G}N_A}\sum_{\mathbf{G}\in \mathcal{G}} |s(\mathbf{G})|^2,
\end{equation}
where $O_{ns}$ is the non-scaled order parameter adopted in the previous study
and $O_{max}$ and $O_{\alpha}$ are scaling parameters.
The set $\mathcal{G}$ is composed of the $N_\mathcal{G}$ shortest reciprocal lattice vectors for
the crystalline order.
$s(\mathbf{G}) = \sum_j \exp(i\mathbf{G}\cdot\mathbf{R}_j)$
is the structure factor where $\mathbf{R}_j$ is the $j$-th atomic position; the sum 
is over $N_A$ atomic positions.
For MgO, only the Mg atom positions were considered in the calculation of $s(\mathbf{G})$.
In the principal calculation condition, which
included 32 Mg and 32 O atoms in a cubic cell,
$O$ is rescaled within $-3/2 < O < 3/2$
with $O_{max} = 32$ and $O_\alpha = 2$. In addition, $N_\mathcal{G} = 8$
for this condition.

\section{The calculation conditions}

The functional form of the model inter-atomic potentials $U_M$ used in this study was the
Born-Huggins-Mayer short-range repulsion potential
with a Morse attractive potential, namely
\begin{equation}
\label{eqn:potfunc}
U_M = \sum_{i<j} \left[ f_0 b_{i,j}\exp\left(\frac{a_{i,j}-r_{i,j}}{b_{i,j}}\right)
    + d_{i,j}\left[\exp\left(-2\beta_{i,j}(r_{i,j} - r^0_{i,j})\right) - 2\exp\left(-\beta_{i,j}(r_{i,j} - r^0_{i,j})\right)\right] \right],
\end{equation}
where $r_{i,j}$ is the distance between atoms $i$ and $j$.
The others, i.e. $a$, $b$, $\beta$, $d$, and $r^0$, are parameters that depend on atomic species, and
are symmetric with respect to the exchange of their indices.
The accurate inter-atomic potential, $U_A$, was obtained by
a density functional calculation
with a plane wave basis set and pseudo-potentials,
which was performed by an extended version of the
program package TAPP.\cite{YTWS96}
The majority of the calculations were performed with
PBE-type XC potential.\cite{PBE96,PBE97}
In others,
PBEsol-type\cite{PRCVSCZB08,PRCVSCZB09}, CAPZ-type\cite{PZ81,CA80}, or PW91-type\cite{GGAII} XC potentials were used.
The principal cell size, which contained 64 atoms in total,
had been used in the previous first-principles study\cite{Alfe05}
to evaluate latent heat and volume change during melting.
Only the $\Gamma$ point was sampled in the first Brillouin zone.
To keep the effective cut-off energy of the basis set constant
against volume change, the method proposed by
Bernasconi et al.\cite{BCFSTP95} was applied; we have
$E_0 = 42.25$ Ry, $A = 80.0$ Ry, and $\sigma = 0.8$ Ry.
The total number of plane waves in the basis set was $23439$.
The effective cut-off $E_0 = 42.25$ Ry is rather higher than that in the
previous first-principles calculation.\cite{Alfe05}
When 4 irreducible $k$-points were sampled in a downfolding iteration,
the changes in the $T_m$, $\Delta H$, and $\Delta V$
compared with the result by $\Gamma$ point sampling
were $10$ K, $2$ kJ$\cdot$mol$^{-1}$, and $-0.08$ \AA$^3$, respectively.

The {\em MOMT} ensemble MD simulations\cite{Yoshimoto06}
were performed using\cite{HOE96,NNK97,OO04bc}
an isobaric-isothermal ensemble MD algorithm.\cite{Stern04}
The reference temperature, $T_0$, for MD was
$2750$, $4350$, $5800$, and $8000$ K for
$P=0$, $17$, $47$ and $135.6$ GPa, respectively.
The parameters of the thermostat
and the barostat, $\omega_\zeta$ and $\omega_\eta$
have values $1.0\times10^{-3}$ and $5.0\times10^{-5}$ in atomic units, respectively.
The time step of the simulation was $5$ a.u.
In the Wang-Landau algorithm that generates the multicanonical weight,
a factor, $f$, was added every $100$ MD steps and the flatness of its histogram
was checked every $30,000$ MD steps.
The mesh spacings for the partial entropy of the system, \cite{OO04bc,Yoshimoto06}
$\delta\tilde{S} = \tilde{S} - U/T_0$,
actually generated in the algorithm were
$0.1$ a.u. for $U$ and $0.1$ for $O$, respectively.
(The unit of the factor and $\delta \tilde{S}$ is $k_B$.)
The maximum value of $\delta \tilde{S}$ was set at zero.
For $P = 0$ GPa and $P = 17$ GPa, the minimum was kept larger than $-40 + 2f$.
For $P = 47$ GPa and $P = 135.6$ GPa, it was kept larger than $-60 + 2f$.
The same definition was
adopted for the flatness of the histogram as employed in the previous study\cite{Yoshimoto06}.
The algorithm was continued until a {\em flat} histogram was obtained
with $f = 1/128$.
At this stage, the multicanonical weight
for the production run was obtained.
The number of MD steps for the production run was $3\times 10^7$.
This was enough to obtain a physically sound smooth temperature
dependence for both liquid and crystalline phases by re-weighting.

The procedure for thermodynamic downfolding was as follows.

To begin, the definition of the target function $L$ for
downfolding was
\begin{equation}
L = \sum_{X_i\in \mathcal{Q}}\left[ \left(\Delta U\left(X_i\right) - \langle\Delta U\rangle\right)^2 + w \left( 3\left.\frac{\partial U_M}{\partial V}\right\vert_{X_i}V_i - 3\left.\frac{\partial U_A}{\partial V}\right\vert_{X_i}V_i\right)^2\right],
\end{equation}
where $\mathcal{Q}$ is a down-sampled multicanonical ensemble.
$X_i$ and $V_i$ represent an atomic configuration
and its corresponding cell volume, respectively.
$\Delta U = U_M - U_A$ is the difference between model and accurate
inter-atomic potential energy. In addition,
\[
\langle\Delta U\rangle = \frac{1}{\vert\mathcal{Q}\vert}\sum_{X_k\in {\cal Q}} \Delta U(X_k).
\]
Atomic units are employed throughout here. $w = 0.02$ is a parameter.
Thus, the present definition of $L$ includes a stress term so that the current study can treat pressure dependence.

As in the previous study on silicon, $X_i$ was considered
in down-sampling for $\mathcal{Q}$ only if
\begin{equation}
\max_O \tilde{S}(U(X_i),O) - \tilde{S}(U(X_i),O(X_i)) \le 7 k_B,
\end{equation}
where $\tilde{S}$ is the estimated entropy by the {\em MOMT} ensemble.
This is because $X_i$ does not otherwise contribute to thermal averages.

At the beginning of the downfolding iteration series,
the effect of thermal electronic excitation was not considered.
Downfolding was performed only under zero pressure.
This stage dealt with generating an initial estimate.

For the first iteration of downfolding, the model inter-atomic potential used
was taken from ref. \citen{Kawamura88}. ($U_m^0$)
The functional form of this potential comprises the Ewald term plus a
Born-Huggins-Mayer term.
After the first iteration, it was found that
the present form of model potential had the same effectiveness as
that with the Ewald term.
Because calculations involving the Ewald term are computationally intensive, the subsequent
calculations were performed with the present form of the potential.

After the first iteration, two further iterations of downfolding were
performed to obtain an initial estimate ($U_m^2$)
with which production iterations can proceed.
In these iterations, thermal electronic excitation effects were considered.

For the first-principles calculations, the number of sampled configurations in each downfolding is set at 500, while
the number of parameters in the present model inter-atomic potential is 9.
In our previous study on silicon which sampled 1000 configurations,\cite{Yoshimoto06}
this number was 16. Therefore,
a smaller sampling set was appropriate.

The first production iteration, S1, was performed under zero pressure.
This iteration produced model potential $U_M^1$ from $U_M^0 = U_m^2$. 

The second production iteration, S2, was performed under
$P$ = $0$, $17$, $47$, and $135.6$ GPa.
In this iteration, multicanonical simulations
were performed with the $U_M^1$ model potential to generate
corresponding down-sampled simulation runs, $\mathcal{Q}^1(P)$, for downfolding.
From the simulation, physical quantities were also obtained.
Using $\mathcal{Q}^1(P)$, downfolding was performed to
obtain $U_M^2(P)$ model potentials for each pressure.

The third production iteration, S3, was performed with pressures
$P$ = $0$, $17$, $47$, and $135.6$ GPa.
In this iteration, multicanonical simulations for each pressure
were performed with the $U_M^2(P)$ model potentials
and physical quantities were obtained from the simulations.

In addition to these main iterations, a branch iteration, \^S3,
with PBEsol was performed
with the same pressure settings: $P$ = $0$, $17$, $47$, and $135.6$ GPa.
To begin, downfolding was performed from
$\mathcal{Q}^1(P)$ in S2 to obtain model potentials $\hat{U}_M^3(P)$.
Then, multicanonical simulations were performed using $\hat{U}_M^3(P)$
to obtain physical quantities.
In addition, a further downfolding was performed at $P$ = $0$ to
obtain $\hat{U}_M^{4}(P=0)$.

Using $\hat{U}_M^{4}(P=0)$, a checking iteration, \^S4, was performed.
From a multicanonical simulation, physical quantities
were obtained to be used as a comparison against results with $\hat{U}_M^{3}(P=0)$.

In addition to PBEsol, another branch iteration, \~S3, with CAPZ
was performed with $P$ = $0$ GPa.
At first, downfolding was performed from
$\mathcal{Q}^1(P=0)$ in S2 to obtain $\tilde{U}_M^3(P=0)$ model potentials.
Subsequently, multicanonical simulations were performed using $\tilde{U}_M^3(P=0)$
to generate a down-sampled simulation run, $\tilde{\mathcal{Q}}^2(P=0)$.
For this branch with CAPZ, the appropriate $T_0$ was $3400$ K.
With $\tilde{\mathcal{Q}}^2(P=0)$, a further downfolding
was performed to obtain $\tilde{U}_M^{4}(P=0)$.

Using $\tilde{U}_M^{4}(P=0)$, a production iteration, \~S4, was performed.
A multicanonical simulation was performed to obtain physical quantities.
By comparing physical quantities of this iteration to previous values, it was found that this iteration is sufficient to obtain results that converged.

As for the computation of the above procedure,
the required computational cost for one iteration of
downfolding and the following
{\em MOMT} simulation is commented in Appendix \ref{sec:apxcompcost}.

\section{Calculational checks }
\label{sec:checks}

First, the effect of thermal electronic excitation was checked at $P=0$ GPa,
by producing a downfolded potential starting from an $U_M^0$
in the absence of the effect. This procedure constituted one iteration of downfolding.
Consequently, $U_M^1$ is the corresponding downfolded potential
with the effect.
Changes in $T_m$, $\Delta H$ and $\Delta V$ were
$150$ K, $-6$ kJ$\cdot$mol$^{-1}$, and $-0.17$ \AA$^3$, respectively.
The thermal electronic excitation clearly decreases the
melting point. Because the energy gap in the electronic structure for the
liquid state vanishes, the decrease in the free energy by 
thermal electronic excitation is significant for the liquid state.
Therefore, it is natural to obtain a lower melting point when
the effect is considered.

Second, the simulation cell-size dependence was studied utilizing two methods.

The first method performs a {\em MOMT} simulation
with a 128-atom fcc cell and a
216-atom cubic cell using a model potential obtained by the 64 atom cubic cell.
The result is shown in Table \ref{tab:natmdep}.
For simulations with larger cells, some techniques were introduced that have been described in
detail in the appendix \ref{sec:apxtech}.
The model potential used was $U_M^2(P=0)$, $\hat{U}_M^3(P=0)$,
and $\tilde{U}_M^4(P=0)$.
From the result, the simulation based on a 64-atom cubic cell
seems to over-estimate $T_m$ by $150$--$200$ K.
The errors in $\Delta V$ and $\Delta H$ were
$\sim 0.5$ \AA$^3$ and $\sim 5$ kJ$\cdot$mol$^{-1}$, respectively.

The decrease in the melting point is understandable because the
larger simulation cell probably enables
the liquid state to take more configurations compared with the crystalline state. Therefore, the relative free energy
of the liquid state should decrease in a larger simulation cell.
This relative decrease in the free energy prompts a lower melting point.

\begin{table}
\caption{\label{tab:natmdep} Simulation cell size dependence of melting
 under $P=0$ GPa.
 A cubic cell containing 64 atoms, a fcc cell containing 128 atoms, and
 a cubic cell containing 216 atoms are compared.
 $U_M^2(P=0)$, $\hat{U}_M^3(P=0)$, and $\tilde{U}_M^4(P=0)$
 model inter-atomic potentials
 were used for PBE, PBEsol, and CAPZ simulations, respectively.
 $P$, $T_m$, $V_x$, $V_l$, $\Delta V$, $\Delta H$, and $\Delta S$ are
 pressure, melting point,
 volume per atom for crystalline state at $T_m$,
 volume per atom for liquid state at $T_m$,
 volume change per atom during melting,
 latent heat, and entropy change per atom during melting, respectively.}
\begin{center}
\begin{tabular}{crrrrrrl}
\hline
atom & \multicolumn{1}{c}{$T_m$}& \multicolumn{1}{c}{$V_x$} & \multicolumn{1}{c}{$V_l$} & \multicolumn{1}{c}{$\Delta V$}& \multicolumn{1}{c}{$\Delta H$}& \multicolumn{1}{c}{$\Delta S$} \\
 &\multicolumn{1}{c}{K} & \multicolumn{1}{c}{\AA$^3$} & \multicolumn{1}{c}{\AA$^3$} & \multicolumn{1}{c}{\AA$^3$} & \multicolumn{1}{c}{kJ$\cdot$mol$^{-1}$} & \multicolumn{1}{c}{$k_B$} \\
 \hline
 64 & 2975 & 10.93 & 14.44 & 3.52 &  85 & 1.7 & PBE \\
128 & 2850 & 10.77 & 14.50 & 3.73 &  86 & 1.8 & PBE \\
216 & 2820 & 10.78 & 14.00 & 3.22 &  82 & 1.7 & PBE \\
\hline
 64 & 3230 & 10.66 & 13.67 & 3.01 &  87 & 1.6 & PBEsol \\
128 & 3120 & 10.57 & 13.80 & 3.23 &  89 & 1.7 & PBEsol \\
216 & 3060 & 10.52 & 13.26 & 2.74 &  86 & 1.7 & PBEsol \\
\hline
 64 & 3460 & 10.43 & 13.26 & 2.84 &  94 & 1.6 & CAPZ \\
128 & 3340 & 10.31 & 13.28 & 2.96 &  93 & 1.7 & CAPZ \\
216 & 3270 & 10.29 & 12.82 & 2.53 &  88 & 1.6 & CAPZ \\
\hline
\end{tabular}
\end{center}
\end{table}

\begin{figure}
  \begin{center}
  \includegraphics{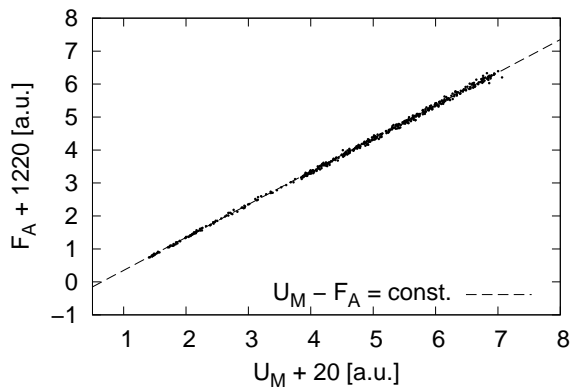}
  \end{center}
  \caption{\label{fig:fcc128-uvsf} Model inter-atomic potential energy $U_M^2(P=0)$ is compared with the first-principles free energy ($F_A$) calculated with 64 Mg and 64 O atoms contained in a fcc cell. The $\Gamma$ point was sampled in the first Brillouin zone. Each dot represents a down sampled configuration in the multicanonical ensemble. 500 configurations were calculated.}
\end{figure}

The second method used to perform calculational checks involves generating a down-sampled multicanonical ensemble
with a model potential for a larger cell and 
comparing this potential to the accurate potential for the same ensemble.
Using a down-sampled simulation run with a 128 atom fcc cell,
the $U_M^2(P=0)$ model potential were compared with the first-principles potential.
The comparison is shown in Fig. \ref{fig:fcc128-uvsf}.
The dashed line in the figure is a guide for the correspondence
between the two potentials. In this
figure, the model potential
and the first-principles calculation agree with each other very well. This means that the $U_M^2(P=0)$ model potential
determined in the 64-atom cubic cell had converged satisfactorily.

\section{Comparison between exchange-correlation potentials}

The present paper compares results using different
exchange-correlation (XC) potentials. However, to compare them,
the residual of thermodynamic downfolding has to be smaller than
the difference due to the XC potentials.
To confirm this, first-principles calculations using
different XC potentials were performed
on the down-sampled multicanonical simulation runs, $\mathcal{Q}^1(P=0)$.
The XC potentials used were
PBE, PBEsol, CAPZ, and PW91.
This comparison on the set $\mathcal{Q}^1(P=0)$ has thermodynamic meaning
because the set represents the thermodynamics of the system.

\begin{figure}
  \begin{center}
  \includegraphics{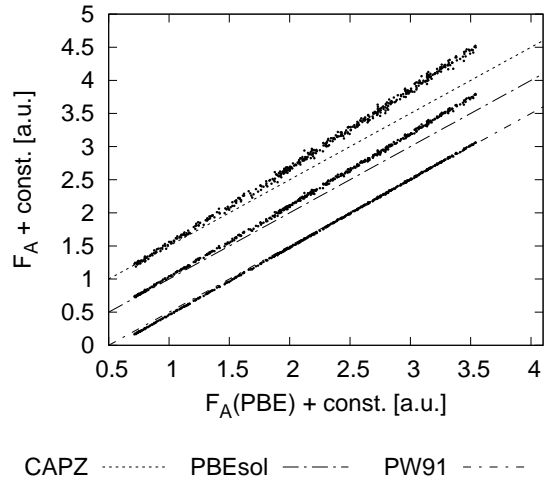}
  \end{center}
  \caption{\label{fig:compXCdirect} Comparison of PBE XC potential with CAPZ, PBEsol and PW91 XC potentials on the multicanonical simulation run, $\mathcal{Q}^1(P=0)$. Each axis displays the free energy of the system.}
\end{figure}

\begin{figure}
  \begin{center}
  \includegraphics{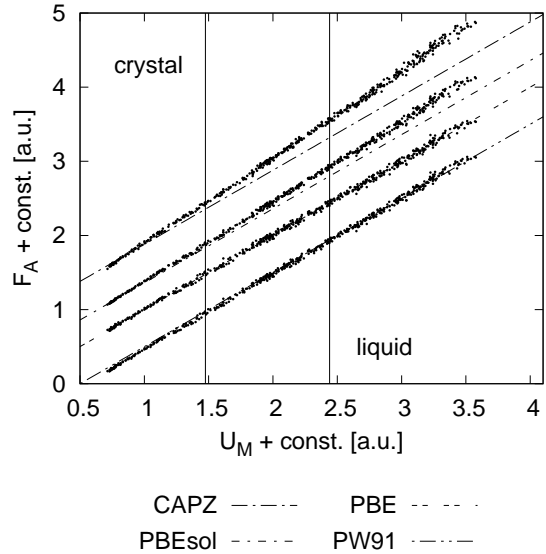}
  \end{center}
  \caption{\label{fig:compXC} Comparison of $U_M^2(P=0)$ with first-principles calculations using CAPZ, PBE, PBEsol and PW91 XC potentials on the multicanonical simulation run, $\mathcal{Q}^1(P=0)$. Each axis displays the free energy of the system. The left(right) vertical line shows the potential energy of crystalline(liquid) state at the melting point.}
\end{figure}

The results are summarized in Figs. \ref{fig:compXCdirect}
and \ref{fig:compXC}. In these figures, there are groups of dots.
Each dot in a group is a member in $\mathcal{Q}^1(P=0)$.
There is a marked line immediately below each group.
The marking indicates a corresponding XC potential found in the legend below the figure.
Each line also determines a relation $x = y + \textrm{const.}$,
where $x$ ($y$) is the vertical (horizontal) axis.

Figure \ref{fig:compXCdirect} shows the comparison between
the calculation using PBE and the calculations using the other XC potentials.
The PW91 curve agrees very well with that for PBE.
Because PBE was developed as a simplified
version of PW91 and was expected to be close to it,
this agreement is considered normal.
The PBEsol curve, however, clearly has a steeper gradient than unity.
This trend is even more significant for the CAPZ curve.

Figure \ref{fig:compXC} shows the comparison between $U_M^2(P=0)$
and the first-principles calculations.
The dots for PBE completely follow the line with no trend.
This means $U_M^2(P=0)$ was appropriately generated.
The spread of dots around the line is because of the residual in
downfolding.
Points associated with PBEsol and CAPZ have steeper gradient than unity in this
comparison as in Fig. \ref{fig:compXCdirect}.
These trends are distinct compared with the spread of points.
This means that the form of the present model inter-atomic potential can distinguish between
PBE, PBEsol and CAPZ.
The dots for PW91, however, follow the line also with no
distinct trend. This means the form of the present model potential 
can not distinguish any difference between PW91 and PBE if such differences existed.

In consequence, the effectiveness of the present method has been verified.
Moreover, this result proves that the present approach, which falls in principle in the category of {\em ab}-{\em initio} methods,
has in practice the capability of an {\em ab}-{\em initio} method.

\section{Melting at $P=0$GPa}
The result at $P=0$ GPa is summarized and compared with
existing theoretical and experimental results in Table \ref{tab:P0summl}.
The table also includes the present results simulated in a 216-atom cell.
The correspondence between the 64- and 216-atom cells has been discussed in \S \ref{sec:checks}.

From theoretical results given by Alf\`e,
the generalized gradient approximation (GGA) produced a significantly
lower melting point compared with the experimental value.
In contrast, the local density approximation (LDA) used in his study produced
a closer melting point.
However, melting points
obtained with PBE or CAPZ in the 64-atom cell rose
by $400$--$450$ K more than that with either GGA or LDA established by Alf\`e.
A partial reason for this
increase is probably because of the effect of the number of the atoms
in the cell, because the present results in 216-atom cell were higher by only
$200$--$300$ K over those by Alf\`e.
(Alf\`e estimated the errors to be $50$ K and $100$ K for LDA and GGA,
respectively.)

However, even if the effect of the number of the atoms
is taken into account,
the present melting point with CAPZ, which is a typical LDA,
was not so significantly better than
that with PBE, which is a typical GGA, in contradistinction to results reported by Alf\`e.
The present melting point with PBEsol was higher by $\sim 250$ K
than that with PBE and in 216 atom cell case was closer to the experimental value.

In addition, the present results for latent heat were
rather close to recent experimental indications
(JANAF\cite{CDDFMS85} and Howard\cite{Howald92}).
This is again in contrast to the LDA result by Alf\`e, which was by a wide margin
larger than experimental indications.
The present values of $\Delta H$ are arranged in ascending order as
$\Delta H$(experimental) $<$ $\Delta H$(PBE) $<$ $\Delta H$(PBEsol) $<$
$\Delta H$(CAPZ).
This ordering is consistent with the mean errors in atomization energy using
these XC potentials\cite{PRCVSCZB08,PRCVSCZB09}.
In a related matter, Tangney and Scandolo
obtained the energy difference between the liquid and crystal
as a function of temperature when they studied the melting slope\cite{TS09}.
The difference was $\sim 79$ kJ$\cdot$mol$^{-1}$ in the range of
$\sim 2950$ K $< T <$ $\sim 3250$ K. Although they did not give
a concluding number for $T_m$, their result means
$\Delta H \sim 79$ kJ$\cdot$mol$^{-1}$. Their simulation cell contained
64 atoms and they used LDA.
The difference between their result and $\Delta H$(CAPZ) is understandable
because they expected error of 10\% and they does not seem to have included
the thermal electronic excitation effect. The effect
caused $6$kJ$\cdot$mol$^{-1}$ change in this study. (\S \ref{sec:checks})

\begin{table}
\caption{\label{tab:P0summl} Summary of results at $P = 0$ GPa compared with those by existing theoretical and experimental works. $T_m$, $\Delta V$, $\Delta H$, and $\Delta S$ are melting point, volume change per atom during melting, latent heat, and entropy change per atom during melting, respectively. For present results, the 216-atom cell results are also shown in parentheses. (See \S \ref{sec:checks} for their details.)}
\begin{center}
\begin{tabular}{lrrrr}
  \hline
  & \multicolumn{1}{c}{$T_m$}& \multicolumn{1}{c}{$\Delta V$}& \multicolumn{1}{c}{$\Delta H$}& \multicolumn{1}{c}{$\Delta S$} \\
  & \multicolumn{1}{c}{K} & \multicolumn{1}{c}{\AA$^3$} & \multicolumn{1}{c}{kJ$\cdot$mol$^{-1}$} & \multicolumn{1}{c}{$k_B$} \\
  \hline
  Present: PBE     & 2975(2820) & 3.52(3.22) &  85(82) & 1.7(1.7) \\
  Present: CAPZ    & 3460(3270) & 2.84(2.53) &  94(88) & 1.6(1.6) \\
  Present: PBEsol  & 3230(3060) & 3.01(2.74) &  87(86) & 1.6(1.7) \\
  Alf\`e: GGA\cite{Alfe05} & 2533 & \\
  Alf\`e: LDA\cite{Alfe05} & 3070 & 3.08 & 112 & 2.19 \\
  \hline
  JANAF\cite{CDDFMS85} & 3100 &     & 78   & 1.5 \\
  Howard\cite{Howald92} &     &     & 74   &    \\
  \hline
\end{tabular}
\end{center}
\end{table}

Possible reasons for the discrepancy in melting point between
the present result with CAPZ and that with LDA by Alf\`e are as follows.
(1) The present approximation for thermal electronic excitations
may be insufficient.
(2) In the two phase simulation by Alf\'e,
the simulation cell was fixed. ($NVE$ simulation)
This might cause an unintended shift of $T_m$.
To study further this discrepancy in melting point,
a thermodynamic integration approach
with a first-principles calculation starting from the present model
potential will be appropriate.
The parameters of the model potentials are listed
in Tables \ref{tab:um1param}, \ref{tab:um2param}, \ref{tab:em2param},
\ref{tab:hum3param}, \ref{tab:hem3param}, and \ref{tab:hum4param}
in Appendix \ref{sec:apxparam}.
These tables report the inter-atomic potentials
for both free energy and energy.

\section{Pressure dependence of melting}
\begin{figure}
  \begin{center}
  \includegraphics{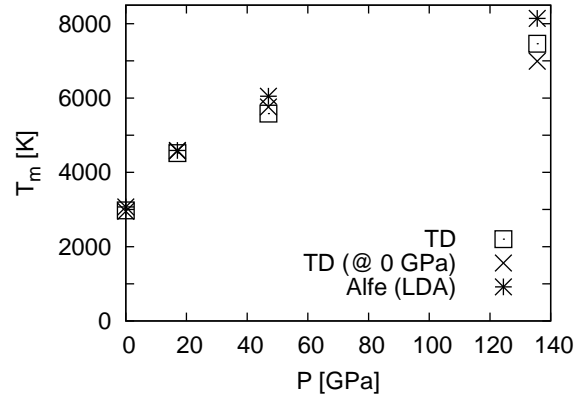}
  \end{center}
  \caption{\label{fig:PvsT} Melting point as a function of pressure. TD, TD(@ 0GPa), and Alf\`e(LDA) are the present result by $U_M^2$, that by $U_M^1$, and the result with LDA by Alf\`e,\cite{Alfe05} respectively.}
\end{figure}

The dependence on pressure of melting point, latent heat, 
and volume change during melting
is shown in Figs. \ref{fig:PvsT}, \ref{fig:PvsdH}, and \ref{fig:PvsVm},
respectively. The numbers are also listed in Table \ref{tab:PQsuml} in
Appendix \ref{sec:apxdetail}.

To obtain the pressure dependence, two methods were tried.
The first is to downfold the model potential at $P=0$ GPa and to
use it for all other pressures. [ TD(@ 0GPa) ]
The second is to downfold the model potential at each pressure. [ TD ]
The latter should yield a better result but the former has the advantage of a lower
computational cost. The convergence of TD in terms of 
downfolding should be sufficient. This is because 
the results obtained by the first generation of downfolding, $U_M^1(P=0)$,
agree well with those by the second one, $U_M^2(P=0)$
(Table \ref{tab:PQsuml}).
This means that one generation from $U_M^1(P), P>0$,
namely, $U_M^2(P), P>0$ is sufficient.

The results obtained by TD(@ 0 GPa) and TD agree rather well
as can be seen in Figs. \ref{fig:PvsT}, \ref{fig:PvsdH}, and \ref{fig:PvsVm}.
In particular, the agreements are rather good for melting points and
for volume changes during melting.

These results mean that the present approach has
worked remarkably well. If the required accuracy
permits, we can generate the potential at any specific pressure and
use it for other nearby pressures.

Alternatively, the pressure dependence obtained here is
similar to the result of Alf\`e. Namely, the present result also indicates
a similar discrepancy between
the first-principles and experimental results\cite{ZB94,ZF08}.

\begin{figure}
  \begin{center}
  \includegraphics{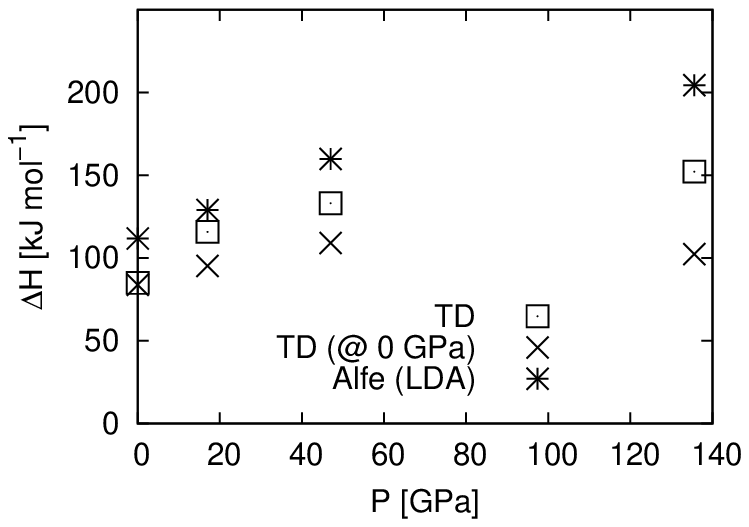}
  \end{center}
  \caption{\label{fig:PvsdH} Latent heat as a function of pressure.  TD, TD(@ 0GPa), and Alf\`e(LDA) are the present result by $U_M^2$, that by $U_M^1$, and the result with LDA by Alf\`e,\cite{Alfe05} respectively.}
\end{figure}

\begin{figure}
  \begin{center}
  \includegraphics{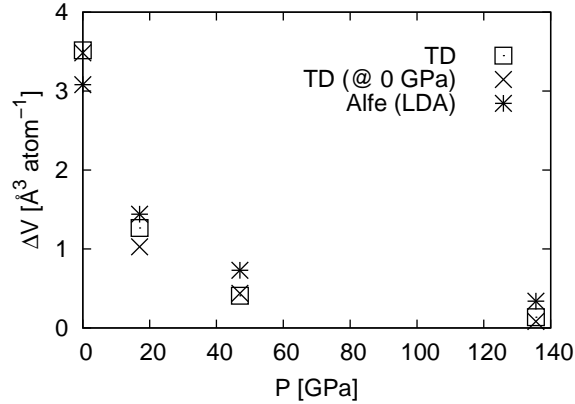}
  \end{center}
  \caption{\label{fig:PvsVm} Volume change during melting as a function of pressure. TD, TD(@ 0GPa), and Alf\`e(LDA) are the present result by $U_M^2$, that by $U_M^1$, and the result with LDA by Alf\`e,\cite{Alfe05} respectively.}
\end{figure}

A possible reason of this discrepancy in the pressure
dependence is an XC potential issue.
Because a claim is made that the new potential PBEsol
is better for condensed materials,
it is worth while to perform the present simulation with PBEsol in addition.

To obtain the result with PBEsol, one generation of downfolding
starting with $U_M^1$ was performed with PBEsol.
The convergence of the generated potential, $\hat{U}_M^3$, is sufficient
because the results obtained with $\hat{U}_M^3(P=0)$ agree well with those
with $\hat{U}_M^4(P=0)$, which represents the second generation.
(Table \ref{tab:PQsuml})

\begin{figure}
  \begin{center}
  \includegraphics{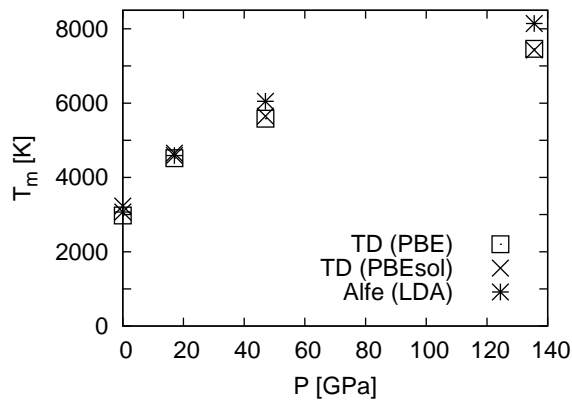}
  \end{center}
  \caption{\label{fig:PvsTsol} Melting point as a function of pressure. TD(PBE), TD(PBEsol), and Alf\`e(LDA) are the present result by $U_M^2$ based on PBE, that by $\hat{U}_M^3$ based on PBEsol, and the result with LDA by Alf\`e,\cite{Alfe05} respectively.}
\end{figure}

\begin{figure}
  \begin{center}
  \includegraphics{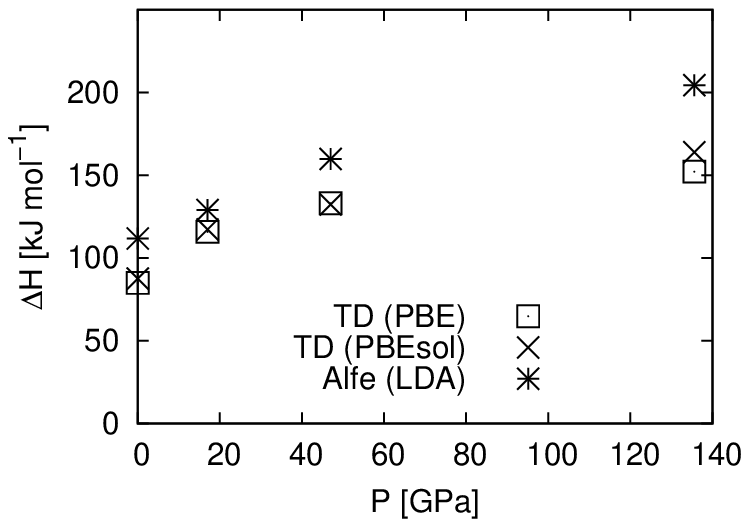}
  \end{center}
  \caption{\label{fig:PvsdHsol} Latent heat as a function of pressure. TD(PBE), TD(PBEsol), and Alf\`e(LDA) are the present result by $U_M^2$ based on PBE, that by $\hat{U}_M^3$ based on PBEsol, and the result with LDA by the Alf\`e,\cite{Alfe05} respectively.}
\end{figure}

\begin{figure}
  \begin{center}
  \includegraphics{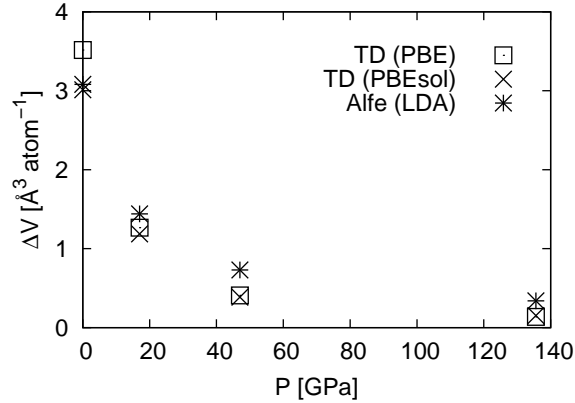}
  \end{center}
  \caption{\label{fig:PvsVmsol} Volume change during melting as a function of pressure. TD(PBE), TD(PBEsol), and Alf\`e(LDA) are the present result by $U_M^2$ based on PBE, that by $\hat{U}_M^3$ based on PBEsol, and the result with LDA by Alf\`e,\cite{Alfe05} respectively.}
\end{figure}

The obtained melting point,
latent heat and volume change during melting are shown
in Figs. \ref{fig:PvsTsol}, \ref{fig:PvsdHsol}, and \ref{fig:PvsVmsol},
respectively. As shown in these figures,
the results with PBE and PBEsol XC potentials are quite close.
In addition, these results are similar to the LDA result by Alf\`e. 
Thus, the discrepancy between theoretical and experimental works is not
resolved even with PBEsol.

To study further the discrepancy arising for the electronic correlation,
we should try a more accurate method such as the diffusion quantum
Monte Carlo method. A reason for doing this is that the melting transition
accompanies the closing of the gap in the electronic structure, and
PBEsol also has a well-known gap issue like CAPZ and PBE.
The present approach can be combined with
such accurate methods because there is no restriction on $U_A$.

In addition to the direct comparison of $T_m$ so far,
the melting slope ($dT_m/dP$) at $P=0$ GPa is calculated using the
Clausius-Clapeyron relation. The results are shown in Table \ref{tab:dTmdP}
with other theoretical and experimental results.
All of the theoretical results in the table were based on density
functional calculations directly (ref. \citen{Alfe05})
or indirectly (ref. \citen{TS09} and \citen{AguadoMadden05}).
The present results are similar to other theoretical results including
the one by Tangney and Scandolo. Thus, a discrepancy
similar to the one in the direct comparison of $T_m$ is
observed in $dT_m/dP$ also.

Besides the theoretical aspect of the discrepancy so far discussed,
we should consider the experimental aspect in addition.
Recently, Adebayo et al. suggested a possible reason for
the discrepancy.\cite{ALMS09}
They studied infrared absorption of MgO at high pressure and temperature
by a MD. They found that the infrared absorption
of crystalline MgO at CO$_2$ laser frequencies increases substantially
with both pressure and temperature.
On the other hand, Zerr and Boehler
observed the abruptness of the absorption changes
with laser intensity to detect the melting.
However, Adebayo et al. claimed
that this abruptness is not necessarily caused by
the melting but can be caused by the nonlinear absorption change
in several way.
Although I regard this as a possible reason,
I also remark that
their opinion seems to have a difficulty to explain why
Zerr and Boehler obtained the established melting temperature under $P=0$ GPa.
If the nonlinear mechanism is absent for this pressure, we
expect some singular point in the melting curve because of the
possible mechanism change. The melting curve given by Zerr and Boehler,
however, does not have such a singular point.

\begin{table}
\caption{\label{tab:dTmdP} The melting slope $dT_m/dP$ [K/GPa] at $P = 0$ GPa. PBE, CAPZ and PBEsol are the present results with the corresponding XC potentials. In the parentheses, the 216 atom results are also given (See \S 6 for their details). Experimental results are marked by asterisks.}
\begin{center}
  \begin{tabular}{cccccccc}
  \hline
  PBE & CAPZ & PBEsol & Alf\`e\cite{Alfe05} & Aguado and & Tangney and & Zerr and & Zhang \\
      &      &        &      & Madden\cite{AguadoMadden05} & Scandolo\cite{TS09} & Boehler\cite{ZB94} & and Fei\cite{ZF08} \\
  \hline
  148   &  126  & 134   & 102 & 125 & $\sim 130$ -- $\sim 150$ & 36$^\ast$ & 221$^\ast$\\
  (133) & (113) & (118) &     &     &  &  & \\
  \hline
\end{tabular}
\end{center}
\end{table}

\section{Conclusion}

In conclusion, melting of MgO has been successfully simulated with
a combination of a multicanonical ensemble method and first-principles
calculation. To take into account thermal excitations of electrons
within the framework of the multicanonical simulation, an approximation
to incorporate the effect into a model inter-atomic potential has been introduced.
The present study used a rather simple
two-body model inter-atomic potential in thermodynamic downfolding.
Significantly, the Ewald term was not contained in the potential.
Nevertheless, thermodynamic downfolding could distinguish
differences due to the
XC potentials used in a first-principles calculations.

Under 0 GPa, the present method was performed separately with PBE, PBEsol, and CAPZ
XC potentials.
Between them, PBEsol seems to give 
a melting point closest to experimental values.
The present values for latent heat using these XC potentials were
rather close to recent experimental results.
This is in contrast to the LDA result of Alf\`e, which was far
larger than those suggested experimentally.
Of the XC potentials used, PBE values came closest
to experiments. This was to be expected
from the mean errors in atomization energy for these XC
potentials.

To obtain the pressure dependence, two methods were tried.
The first was to downfold the model potential at $P=0$ GPa and to
use results for all other pressures.
The second was to downfold the model potential at each pressure.
Results showed that these two methods agreed well one with the other.
This suggests that the present approach worked remarkably well.

The obtained pressure dependence was similar to the previous study by Alf\`e.
PBEsol, which is a revised parameterization of PBE
and claims better performance for condensed systems,
did not change this dependence. Therefore, the discrepancy between
first-principles and experimental studies unfortunately
still remains in the melting curve of MgO.
The alternative comparison of the melting slope at $P=0$ GPa
also shows a similar discrepancy, which was also claimed by Tangney and
Scandolo.

\begin{acknowledgments}
This work is partially supported by a Grant-in-Aid for 
Young Scientists (B) of the Ministry of Education, Culture,
Sports, Science, and Technology (MEXT), Japan, by a Grant-in-Aid for Scientific
Research in Priority Areas ``Development of New Quantum Simulators and
Quantum Design'' (No.17064004) of the MEXT, Japan, and by the Next
Generation Super Computing Project, Nanoscience Program, MEXT, Japan.
The computation in this work had been done using
the facilities of the Supercomputer Center, Institute for Solid State
Physics, University of Tokyo, the facilities of Supercomputing Division,
Information Technology Center, The University of Tokyo,
and a facility of Institute for Molecular Science, National
Institute of Natural Science, Japan.
\end{acknowledgments}

\appendix

\section{Technical improvements for larger simulation cells}
\label{sec:apxtech}
For larger simulation cells, two additional technical improvements were
introduced.

First, a minor control affecting the Wang-Landau factor was set up.

In large simulation cells, nearly perfect crystalline orderings become
possible. These are characterized by the reciprocal lattice vectors
whose lengths are nearly the same as those for perfect crystalline ordering.
These nearly perfect structures are irrelevant in the re-weighting to
calculate physical quantities because they are observed in the
small entropy area of (energy, order parameter) space in current system sizes.
However, these become traps during simulations.

These traps can be avoided in the learning process of the Wang-Landau
algorithm by imposing a penalty on the Wang-Landau factor
when these are accessed.
Specifically, the factor was multiplied by $A_P = 32$ when
\begin{equation}
\frac{1}{N_{A}^p}\sqrt{\frac{1}{|\mathcal{G}_{\textrm{avd}}|}\sum_{\mathbf{G}\in\mathcal{G}_{\textrm{avd}}}|s(\mathbf{G})|^{2p}}
\end{equation}
becomes larger than a specified threshold.
The number of reciprocal vectors $\mathcal{G}_{\textrm{avd}}$
needed to invoke this ¡Èavoidance¡É was 96 for the 128-atom fcc cell and 144
for 216-atom cubic cell, respectively.
The exponent $p = 3$ sharpens the
discrimination of the quasi-order.
The threshold was $0.07$ for the 128 -atom fcc cell and $0.02$ for the 216-atom cubic
cell.
In production runs, this ``avoidance'' is disabled so as to keep
physical quantities.

For production runs, a related improvement is also available.
In the learning process, we record a histogram for the application
of this penalty. The area in (energy, order parameter) space
where this histogram is above a suitable threshold
can be regarded as the trapping area.
Exploiting this fact, we define an appropriate function that is positive on the trapping area
and zero otherwise.
We can make the production run avoid the area by adding
this function to the multicanonical weight.
An example of the function is the logarithm of the
histogram itself. When the height of the function is 3$k_B$,
the probability of observing the system in the area is reduced to $1/\exp(3)$.

Second, for the 216-atom cubic cell,
the definition of the scaled order parameter changes to
\begin{equation}
O = 3\left(\frac{O_{ns}}{O_{max}}\right)^{\alpha} - \frac{3}{2},
\end{equation}
where $O_{max}$ and $\alpha$
are $108$ and $\log(2/3)/\log(1/2)$, respectively.
The number of shortest reciprocal lattice vectors was 
32 for this cell. This number is large
because the atomic arrangement of perfect crystalline order is not
unique in this cell. Nevertheless, the number of simultaneously
active reciprocal vectors was 8 and consequently normalization by
the set size $|\mathcal{G}|$ in the definition of $O_{ns}$ was
decreased to 8 for this case.
Also the effective range of $O$ was from $-3/2$ to $3/2$ by this definition.

\section{Details of the results}\label{sec:apxdetail}

In Table \ref{tab:PQsuml}, details are listed of the results from
$U_M^1$, $U_M^2$, $\hat{U}_M^3$, $\hat{U}_M^4$ and $\tilde{U}_M^4$.
$U_M^1$ and $U_M^2$ used PBE, $\hat{U}_M^3$ and $\hat{U}_M^4$ used PBEsol and
$\tilde{U}_M^4$ used CAPZ.

\begin{fulltable}
\caption{\label{tab:PQsuml} Summary of the results for a 64 atom cubic cell.
 $P$, $T_m$, $V_x$, $V_l$, $\Delta V$, $\Delta H$, and $\Delta S$ are
 pressure, melting point,
 volume per atom for crystalline state at $T_m$,
 volume per atom for liquid state at $T_m$,
 volume change per atom during melting,
 latent heat, and entropy change per atom during melting, respectively.
 The far left column displays the model inter-atomic potential used while
 the far right column displays the XC potential used.}
\begin{center}
\begin{tabular}{lrrrrrrrl}
 \hline
 & \multicolumn{1}{c}{$P$} &\multicolumn{1}{c}{$T_m$}& \multicolumn{1}{c}{$V_x$} & \multicolumn{1}{c}{$V_l$} & \multicolumn{1}{c}{$\Delta V$}& \multicolumn{1}{c}{$\Delta H$}& \multicolumn{1}{c}{$\Delta S$}& \\
 &\multicolumn{1}{c}{GPa} &\multicolumn{1}{c}{K} & \multicolumn{1}{c}{\AA$^3$} & \multicolumn{1}{c}{\AA$^3$} & \multicolumn{1}{c}{\AA$^3$} & \multicolumn{1}{c}{kJ$\cdot$mol$^{-1}$} & \multicolumn{1}{c}{$k_B$} & \\
 \hline
 $U_M^1(P=0)$ &   0   & 2950 & 10.94 & 14.42 & 3.49 &  84 & 1.7 & PBE\\
 $U_M^1(P=0)$ &  17   & 4575 & 10.03 & 11.05 & 1.03 &  95 & 1.3 & PBE\\
 $U_M^1(P=0)$ &  47   & 5760 &  8.74 &  9.18 & 0.44 & 109 & 1.1 & PBE\\
 $U_M^1(P=0)$ & 135.6 & 6990 &  6.94 &  7.03 & 0.08 & 102 & 0.9 & PBE\\
\hline
 $U_M^2(P=0)$ &   0   & 2975 & 10.93 & 14.44 & 3.52 &  85 & 1.7 & PBE\\
 $U_M^2(P=17)$ &   17   & 4519 &  9.98 & 11.25 & 1.27 & 116 & 1.5 & PBE\\
 $U_M^2(P=47)$ &  47   & 5580 &  8.84 &  9.24 & 0.41 & 133 & 1.4 & PBE\\
 $U_M^2(P=135.6)$ & 135.6 & 7460 &  7.12 &  7.25 & 0.14 & 152 & 1.2 & PBE\\
\hline
 $\hat{U}_M^3(P=0)$ &   0   & 3230 & 10.66 & 13.67 & 3.01 &  87 & 1.6 & PBEsol\\
 $\hat{U}_M^3(P=17)$ &  17   & 4655 &  9.77 & 10.96 & 1.19 & 117 & 1.5 & PBEsol\\
 $\hat{U}_M^3(P=47)$ &  47   & 5650 &  8.67 &  9.05 & 0.39 & 132 & 1.4 & PBEsol\\
 $\hat{U}_M^3(P=135.6)$ & 135.6 & 7430 &  6.99 &  7.15 & 0.15 & 164 & 1.3 & PBEsol\\
\hline
 $\hat{U}_M^4(P=0)$ &   0   & 3170 & 10.69 & 14.00 & 3.31 &  85 & 1.6 & PBEsol\\
\hline
 $\tilde{U}_M^4(P=0)$ &   0   & 3460 & 10.43 & 13.26 & 2.84 & 93.9 & 1.6 & CAPZ\\
 \hline
\end{tabular}
\end{center}
\end{fulltable}

\section{Downfolded model inter-atomic potentials}
\label{sec:apxparam}
The downfolded parameters for the model inter-atomic potentials are listed in
Tables \ref{tab:um1param}, \ref{tab:um2param}, \ref{tab:em2param},
\ref{tab:hum3param}, \ref{tab:hem3param}, \ref{tab:hum4param},
and \ref{tab:tum4param}.
In these tables, $U$ and $E$ are the free energy 
and its associated energy inter-atomic potentials, respectively.
Their functional form is presented in the main text (eq. \ref{eqn:potfunc}).
Both potentials had this same functional form. The value of
$f_0$ was $8.4333463\times 10^{-4}$. The parameters not given in
the tables were set to zero.

\begin{fulltable}
\caption{\label{tab:um1param} Downfolded parameters
for the $U_M^1(P=0)$ and its associated $E_M^1(P=0)$ potentials.
All non-dimensionless values are in atomic units.}
\begin{center}
{\footnotesize
\begin{tabular}{cllllllllll}
  \hline
 & \multicolumn{1}{c}{$a_{\textrm{Mg},\textrm{Mg}}$} & \multicolumn{1}{c}{$a_{\textrm{Mg},\textrm{O}}$} & \multicolumn{1}{c}{$a_{\textrm{O},\textrm{O}}$} & \multicolumn{1}{c}{$b_{\textrm{Mg},\textrm{Mg}}$}  & \multicolumn{1}{c}{$b_{\textrm{Mg},\textrm{O}}$} & \multicolumn{1}{c}{$b_{\textrm{O},\textrm{O}}$} & \multicolumn{1}{c}{$d_{\textrm{Mg},\textrm{O}}$} & \multicolumn{1}{c}{$\beta_{\textrm{Mg},\textrm{O}}$} & \multicolumn{1}{c}{$r^0_{\textrm{Mg},\textrm{O}}$} \\
 \hline
 $U$ & 9.037007 &  8.922508 & 9.139069 & 1.030136 & 1.019860 & 1.050079 & 0.440839 & 0.938921 & 2.515958 \\
 $E$ & 9.521568 & 11.222058 & 9.984526 & 1.120906 & 1.096115 & 1.196248 & 5.604914 & 0.894860 & 1.060691 \\
 \hline
\end{tabular}
}
\end{center}
\end{fulltable}

\begin{fulltable}
\caption{\label{tab:um2param} Downfolded parameters
for the $U_M^2(P)$ potentials. $P$ is in GPa.
All non-dimensionless values are in atomic unit.}
\begin{center}
{\footnotesize
\begin{tabular}{lllllllllll}
 \hline
 \multicolumn{1}{c}{$P$} & \multicolumn{1}{c}{$a_{\textrm{Mg},\textrm{Mg}}$} & \multicolumn{1}{c}{$a_{\textrm{Mg},\textrm{O}}$} & \multicolumn{1}{c}{$a_{\textrm{O},\textrm{O}}$} & \multicolumn{1}{c}{$b_{\textrm{Mg},\textrm{Mg}}$}  & \multicolumn{1}{c}{$b_{\textrm{Mg},\textrm{O}}$} & \multicolumn{1}{c}{$b_{\textrm{O},\textrm{O}}$} & \multicolumn{1}{c}{$\beta_{\textrm{Mg},\textrm{O}}$} & \multicolumn{1}{c}{$d_{\textrm{Mg},\textrm{O}}$} & \multicolumn{1}{c}{$r^0_{\textrm{Mg},\textrm{O}}$} \\
 \hline
 $0$     & 9.511351 & 10.285953 & 9.227619 & 1.147506 & 1.032505 & 1.064371 & 2.457641 & 0.941071 & 1.571412\\
 $17$    & 7.805613 & 10.839585 & 7.749410 & 0.758634 & 1.249861 & 0.783224 & 0.757560 & 0.879671 & 2.328365\\
 $47$    & 8.508896 &  9.993460 & 8.096387 & 0.914784 & 1.121834 & 0.867463 & 0.682049 & 0.915604 & 2.309614\\
 $135.6$ & 8.991686 &  9.402277 & 8.207214 & 0.990086 & 1.007392 & 0.873787 & 0.847770 & 0.967281 & 2.113917\\
 \hline
\end{tabular}
}
\end{center}
\end{fulltable}

\begin{fulltable}
\caption{\label{tab:em2param} Downfolded parameters
for the $E_M^2(P)$ potentials associated with $U_M^2(P)$. $P$ is in GPa.
All non-dimensionless values are in atomic unit.}
\begin{center}
{\footnotesize
\begin{tabular}{lllllllllll}
 \hline
 \multicolumn{1}{c}{$P$} & \multicolumn{1}{c}{$a_{\textrm{Mg},\textrm{Mg}}$} & \multicolumn{1}{c}{$a_{\textrm{Mg},\textrm{O}}$} & \multicolumn{1}{c}{$a_{\textrm{O},\textrm{O}}$} & \multicolumn{1}{c}{$b_{\textrm{Mg},\textrm{Mg}}$}  & \multicolumn{1}{c}{$b_{\textrm{Mg},\textrm{O}}$} & \multicolumn{1}{c}{$b_{\textrm{O},\textrm{O}}$} & \multicolumn{1}{c}{$d_{\textrm{Mg},\textrm{O}}$} & \multicolumn{1}{c}{$\beta_{\textrm{Mg},\textrm{O}}$} & \multicolumn{1}{c}{$r^0_{\textrm{Mg},\textrm{O}}$} \\
 \hline
 $0$     &  10.057433 & 10.958293 & 10.093550 & 1.238631 & 1.117503 & 1.220149 & 3.077068 & 0.868479 & 1.364818 \\
 $17$    &   8.586335 &  9.654085 &  8.928195 & 0.904799 & 0.904578 & 0.980405 & 3.180478 & 1.065285 & 1.609773 \\
 $47$    &   8.802583 &  9.027273 &  8.648471 & 0.943127 & 0.832501 & 0.910779 & 2.431115 & 1.128023 & 1.785129 \\
 $135.6$ &   8.834105 &  9.124699 &  8.447506 & 0.940447 & 0.924673 & 0.857453 & 0.997715 & 1.032836 & 2.236116 \\
 \hline
\end{tabular}
}
\end{center}
\end{fulltable}

\begin{fulltable}
\caption{\label{tab:hum3param} Downfolded parameters
for the $\hat{U}_M^3(P)$ potentials based on PBEsol. $P$ is in GPa.
All non-dimensionless values are in atomic unit.}
\begin{center}
{\footnotesize
\begin{tabular}{lllllllllll}
 \hline
 \multicolumn{1}{c}{$P$} & \multicolumn{1}{c}{$a_{\textrm{Mg},\textrm{Mg}}$} & \multicolumn{1}{c}{$a_{\textrm{Mg},\textrm{O}}$} & \multicolumn{1}{c}{$a_{\textrm{O},\textrm{O}}$} & \multicolumn{1}{c}{$b_{\textrm{Mg},\textrm{Mg}}$}  & \multicolumn{1}{c}{$b_{\textrm{Mg},\textrm{O}}$} & \multicolumn{1}{c}{$b_{\textrm{O},\textrm{O}}$} & \multicolumn{1}{c}{$d_{\textrm{Mg},\textrm{O}}$} & \multicolumn{1}{c}{$\beta_{\textrm{Mg},\textrm{O}}$} & \multicolumn{1}{c}{$r^0_{\textrm{Mg},\textrm{O}}$} \\
 \hline
 $0$     &  9.616249 & 10.416043 & 9.381960 & 1.169388 & 1.056588 & 1.104599 & 2.464106 & 0.918641 & 1.526871 \\
 $17$    &  7.809186 & 10.902950 & 7.719037 & 0.760681 & 1.237988 & 0.783028 & 0.878400 & 0.879261 & 2.240094 \\
 $47$    &  8.479070 &  9.991900 & 8.049807 & 0.910601 & 1.120429 & 0.864588 & 0.692440 & 0.917065 & 2.299699 \\
 $135.6$ &  8.946077 &  9.367895 & 8.160640 & 0.984307 & 1.001234 & 0.870410 & 0.852599 & 0.974624 & 2.113341 \\
 \hline
\end{tabular}
}
\end{center}
\end{fulltable}

\begin{fulltable}
\caption{\label{tab:hem3param} Downfolded parameters
for the $\hat{E}_M^3(P)$ potentials associated with $\hat{U}_M^3(P)$.
(based on PBEsol) $P$ is in GPa.
All non-dimensionless values are in atomic unit.}
\begin{center}
{\footnotesize
\begin{tabular}{lllllllllll}
 \hline
 \multicolumn{1}{c}{$P$} & \multicolumn{1}{c}{$a_{\textrm{Mg},\textrm{Mg}}$} & \multicolumn{1}{c}{$a_{\textrm{Mg},\textrm{O}}$} & \multicolumn{1}{c}{$a_{\textrm{O},\textrm{O}}$} & \multicolumn{1}{c}{$b_{\textrm{Mg},\textrm{Mg}}$}  & \multicolumn{1}{c}{$b_{\textrm{Mg},\textrm{O}}$} & \multicolumn{1}{c}{$b_{\textrm{O},\textrm{O}}$} & \multicolumn{1}{c}{$d_{\textrm{Mg},\textrm{O}}$} & \multicolumn{1}{c}{$\beta_{\textrm{Mg},\textrm{O}}$} & \multicolumn{1}{c}{$r^0_{\textrm{Mg},\textrm{O}}$} \\
 \hline
 $0$     & 10.173488 & 11.031455 & 10.223373 & 1.262270 & 1.142935 & 1.252087 & 2.803954 & 0.848087 & 1.375338 \\
 $17$    &  8.583008 &  9.650199 &  8.930708 & 0.905555 & 0.904484 & 0.986814 & 3.202733 & 1.064061 & 1.600048 \\
 $47$    &  8.763038 &  8.978098 &  8.598636 & 0.936597 & 0.828123 & 0.905677 & 2.346935 & 1.132265 & 1.802724 \\
 $135.6$ &  8.779660 &  9.148105 &  8.390658 & 0.932058 & 0.931337 & 0.850587 & 0.969167 & 1.033046 & 2.259133 \\
 \hline
\end{tabular}
}
\end{center}
\end{fulltable}

\begin{fulltable}
\caption{\label{tab:hum4param} Downfolded parameters
for the $\hat{U}_M^4(P=0)$ and its associated $\hat{E}_M^4(P=0)$ potentials.
These were based on PBEsol.
All non-dimensionless values are in atomic unit.}
\begin{center}
{\footnotesize
\begin{tabular}{cllllllllll}
 \hline
 & \multicolumn{1}{c}{$a_{\textrm{Mg},\textrm{Mg}}$} & \multicolumn{1}{c}{$a_{\textrm{Mg},\textrm{O}}$} & \multicolumn{1}{c}{$a_{\textrm{O},\textrm{O}}$} & \multicolumn{1}{c}{$b_{\textrm{Mg},\textrm{Mg}}$}  & \multicolumn{1}{c}{$b_{\textrm{Mg},\textrm{O}}$} & \multicolumn{1}{c}{$b_{\textrm{O},\textrm{O}}$} & \multicolumn{1}{c}{$d_{\textrm{Mg},\textrm{O}}$} & \multicolumn{1}{c}{$\beta_{\textrm{Mg},\textrm{O}}$} & \multicolumn{1}{c}{$r^0_{\textrm{Mg},\textrm{O}}$} \\
 \hline
 $U$ & 8.831134 &  10.237103 & 8.777481 & 0.979987 & 1.019958 & 0.974609 & 2.384864 & 0.970088 & 1.663677 \\
 $E$ & 8.831114 &  10.431142 & 8.779725 & 0.979918 & 1.015092 & 0.975064 & 3.449138 & 0.975032 & 1.474028 \\
 \hline
\end{tabular}
}
\end{center}
\end{fulltable}

\begin{fulltable}
\caption{\label{tab:tum4param} Downfolded parameters
for the $\tilde{U}_M^4(P=0)$ and its associated $\tilde{E}_M^4(P=0)$ potentials.
These were based on CAPZ.
All non-dimensionless values are in atomic unit.}
\begin{center}
{\footnotesize
\begin{tabular}{cllllllllll}
 \hline
 & \multicolumn{1}{c}{$a_{\textrm{Mg},\textrm{Mg}}$} & \multicolumn{1}{c}{$a_{\textrm{Mg},\textrm{O}}$} & \multicolumn{1}{c}{$a_{\textrm{O},\textrm{O}}$} & \multicolumn{1}{c}{$b_{\textrm{Mg},\textrm{Mg}}$}  & \multicolumn{1}{c}{$b_{\textrm{Mg},\textrm{O}}$} & \multicolumn{1}{c}{$b_{\textrm{O},\textrm{O}}$} & \multicolumn{1}{c}{$d_{\textrm{Mg},\textrm{O}}$} & \multicolumn{1}{c}{$\beta_{\textrm{Mg},\textrm{O}}$} & \multicolumn{1}{c}{$r^0_{\textrm{Mg},\textrm{O}}$} \\
 \hline
 $U$ & 9.177385 &  10.329528 & 8.952782 & 1.052954 & 1.042900 & 1.021168 & 2.428146 & 0.939454 & 1.563449 \\
 $E$ & 8.984235 &  9.8872164 & 8.922903 & 1.008925 & 0.949289 & 0.985487 & 3.216059 & 1.014374 & 1.489500 \\
 \hline
\end{tabular}
}
\end{center}
\end{fulltable}

\section{Computational costs}
\label{sec:apxcompcost}
It is difficult to compare computational costs between different theoretical works,
because the details in algorithms, program codes and hardware become important.
In addition, the present study used a variety of computers.
Nevertheless, to give some idea of the required computational costs for the
present study, costs for one iteration of 
downfolding and the following {\em MOMT} simulation is presented here.

In total 605 CPU hours on a SGI Altix 3700Bx2 in the Institute for
Solid State Physics (ISSP) was needed to perform the first-principles calculation of
thermodynamic downfolding in S2 iteration at $P=47$ GPa. By
parallelizing the computation, the ¡Èwall-clock¡É time for the calculation
was 5 hours. The following {\em MOMT} simulation in S3 iteration was
performed by a node of Hitachi SR11000/J1 in ISSP. To obtain a
converged $\tilde{W}$, 20 node hours was used. The production run
consumed 10 node hours.

The wall clock time for the {\em MOMT} simulation is larger than
that for downfolding. However, this may be because TAPP is far more
developed compared with the program performing the {\em MOMT} simulation.


\end{document}